\newif\ifpreprint
\preprinttrue
\newif\ifcomment
\newif\ifnprl   
\newif\ifdraft
\ifpreprint
\documentclass[ALICE,manyauthors]{cernphprep}
\usepackage[comma,square,numbers,sort&compress]{natbib}
\else
\documentclass[10pt,aps,prl,twocolumn,superscriptaddress,altaffilletter,nobibnotes,nofootinbib]{revtex4-1}
\usepackage{graphicx}
\fi 
\usepackage[utf8]{inputenc}
\usepackage[T1]{fontenc}
\usepackage{hyperref}
\usepackage{lineno}
\usepackage{amsmath}
\usepackage{epsfig}
\usepackage{color}
\usepackage[export]{adjustbox}
\usepackage{xspace}
\usepackage[T1]{fontenc}
\usepackage{orcidlink} 

\definecolor{dgreen}{cmyk}{1.,0.,1.,0.1}      
\definecolor{orange}{cmyk}{0.,0.353,1.,0.}    

\newcommand{\com}[1]    {\relax}

\newcommand{ \la }{\langle}
\newcommand{ \ra }{\rangle}

\def\dvers {v8}

\def\PbPb  {\mbox{Pb--Pb}\xspace}

\def\snn   {\mbox{$\sqrt{s_{_{\rm NN}}}$}\xspace}
\newcommand{\Fpsi}[1]{\ensuremath{A^{\rm f}_#1}}
\newcommand{\Fvn}[1]{\ensuremath{M^{\rm f}_#1}}
\newcommand{\trento}{\ensuremath{\rm T_{R}ENTo}\xspace}
\newcommand{\pt}{\ensuremath{p_{\rm T}}\xspace}
\newcommand{\ptpoi}{\ensuremath{p_{\rm T}^{\rm POI}}\xspace}

\newcommand{\vn}{\ensuremath{v_{n}}\xspace}

\newcommand{\vnsq}{\ensuremath{\vn^{2}}\xspace}

\graphicspath{{.img/}}

\begin{document}%
\ifpreprint
\PHyear{2022}
\PHnumber{121}      
\PHdate{07 June}  
\begin{titlepage}
\fi
\title{Observation of flow angle and flow magnitude fluctuations in \PbPb\ collisions at \snn = 5.02 TeV at the CERN Large Hadron Collider}
\ifpreprint
\ShortTitle{Flow angle and flow magnitude fluctuations in \PbPb\ collisions at the LHC}   
\Collaboration{ALICE Collaboration\thanks{See Appendix~\ref{app:collab} for the list of collaboration members}}
\ShortAuthor{ALICE Collaboration} 
\fi
\ifdraft
\begin{center}
\today\\ \color{red}DRAFT \dvers\ \hspace{0.3cm} \$Revision: 5573$\color{white}:$\$\color{black}\vspace{0.3cm}
\end{center}

\fi
\begin{abstract}
This Letter reports on the first measurements of transverse momentum dependent flow angle $\Psi_n$ and flow magnitude $v_n$ fluctuations, determined using new four-particle correlators. The measurements are performed for various centralities in \PbPb collisions at a centre-of-mass energy per nucleon pair of \snn = 5.02 TeV with ALICE at the CERN Large Hadron Collider. Both flow angle and flow magnitude fluctuations are observed in the presented centrality ranges and are strongest in the most central collisions and for a transverse momentum $\pt>2$ GeV/\textit{c}. Comparison with theoretical models, including iEBE-VISHNU, MUSIC, and AMPT, show that the measurements exhibit unique sensitivities to the initial state of heavy-ion collisions. 
\end{abstract}
\ifpreprint
\end{titlepage}                                                                                                                                              
\setcounter{page}{2}
\else
\maketitle
\fi


In ultrarelativistic collisions of heavy ions, such as those at the BNL Relativistic Heavy-Ion Collider (RHIC) and the CERN Large Hadron Collider (LHC), a deconfined state of strongly interacting matter, commonly referred to as quark--gluon plasma (QGP), is predicted to be created under extreme conditions of temperature and energy densities~\cite{Shuryak:1978ij, Shuryak:1980tp}. Many experimental results indicate that a strongly-coupled QGP is formed in heavy-ion collisions~\cite{BRAHMS:2004adc,STAR:2005gfr,PHENIX:2004vcz,PHOBOS:2004zne,Muller:2012zq}. Initial anisotropies of the geometric overlap of the colliding nuclei and spatial inhomogeneities in the energy density drive the collective expansion of the QGP and are transformed through the evolution of the QGP into a momentum anisotropy in the final state~\cite{Ollitrault:1992bk, Voloshin:2008dg, Heinz:2013th}. This momentum anisotropy is characterised by a Fourier expansion of the distribution of the azimuthal angle, $\varphi$, of emitted particles~\cite{Voloshin:1994mz}
\begin{align}
\frac{\mathrm{d}^2N}{\mathrm{d}\pt\mathrm{d}\varphi}= \frac{\mathrm{d}N}{2\pi\mathrm{d}\pt}\left(1+2\sum_{n=1}^\infty v_n(\pt)\cos[n(\varphi-\Psi_n(\pt))]\right),\label{eq:dNdPhi}
\end{align}
where $v_n(\pt)$ and $\Psi_n(\pt)$ correspond to the magnitude and angle, respectively, of the $n^\mathrm{th}$-order harmonic flow vector $\vec{V}_n(\pt) = v_n(\pt)e^{in\Psi_n(\pt)}$. Here the transverse momentum, \pt, dependence of both the flow magnitude and flow angle has been made explicit. The flow vector quantifies the orientation and magnitude of the anisotropic flow, and the flow angle $\Psi_n$ is the angle of the symmetry plane of the $n^\mathrm{th}$-order flow vector.

Typically, the largest flow coefficient is the elliptic flow $v_2$, since it is largely determined by the geometrical overlap of the colliding nuclei. However, fluctuations in the position of the colliding nuclei and in the position of nucleons within the nuclei can induce more complex geometrical shapes, which will give rise to non-zero flow coefficients with $n>2$~\cite{Alver:2010gr,Alver:2010dn,Teaney:2010vd,Luzum:2010sp}, such as the triangular flow $v_3$. Both elliptic and triangular flow coefficients have been measured extensively at RHIC~\cite{STAR:2000ekf,PHENIX:2003qra,STAR:2013qio,PHENIX:2015idk} and the LHC~\cite{Aamodt:2010pa,ALICE:2011ab,Abelev:2014pua,Adam:2016izf,Acharya:2017zfg,ATLAS:2012at,ATLAS:2011ah,Aad:2013xma,Chatrchyan:2012wg,CMS:2012zex,Chatrchyan:2012xq,Acharya:2018ihu}. The comparison to hydrodynamical calculations can constrain the initial conditions of the heavy-ion collisions and the transport properties of the QGP, such as the specific shear viscosity $\eta/s$. These comparisons indicate that the system behaves as a strongly-coupled low-viscosity fluid~\cite{Heinz:2013th,Luzum:2013yya,Shuryak:2014zxa,Song:2017wtw,ALICE:2022wpn}. 

Fluctuations of the flow angle $\Psi_n$ and the flow magnitude $v_n$ have been shown to be present in hydrodynamical models~\cite{Heinz:2013bua,Gardim:2012im}. These fluctuations are possibly due to thermal hydrodynamic fluctuations during the QGP evolution~\cite{Sakai:2020pjw}. The \textit{flow angle fluctuations} (FAF) are the fluctuations of $\Psi_n(\pt)$ determined by a subset of particles at a specific \pt with respect to the symmetry plane determined by the total set of particles, $\Psi_n$. If such fluctuations are present, $\Psi_n(\pt) \neq \Psi_n$. The \textit{flow magnitude fluctuations} (FMF) of the $v_n$ at different \pt can be understood as a decorrelation where $\la v_n(\pt)v_n\ra \neq \sqrt{\la v_n^2(\pt)\ra\la v_n^2\ra}$. 

The determination of QGP properties, such as $\eta/s$, rely on the comparison of theoretical model calculations to experimental data. In order to provide an unbiased extraction of the QGP properties, the models should account for the fluctuations in the flow angle and flow magnitude. Measurements at the LHC have reported the existence of \pt-dependent flow vector fluctuations~\cite{ALICE:2017lyf,CMS:2013bza,Khachatryan:2015oea,Aamodt:2011by}. However, those measurements rely on observables constructed from two-particle correlations, which intrinsically contain contributions from both the FAF and FMF with no way to separate the two experimentally.

In this Letter the FAF and FMF are measured with two new four-particle correlation functions to separate the components of the \pt-dependent fluctuations of the flow vector. 

The FAF is quantified by
\begin{align}
A^\mathrm{f}_n = \frac{\la\la\cos n[\varphi_1^\mathrm{POI}+\varphi_2^\mathrm{POI}-\varphi_3-\varphi_4]\ra\ra}{\la\la\cos n[\varphi_1^\mathrm{POI}+\varphi_2-\varphi_3^\mathrm{POI}-\varphi_4]\ra\ra}&= \frac{\la v_n^2(p_\mathrm{T})~v_n^2 \cos 2n[\Psi_n(p_\mathrm{T})-\Psi_n]\ra}{\la v_n^2(p_\mathrm{T})v_n^2\ra}\nonumber\\
&\simeq \la \cos 2n\left[\Psi_n(\pt)-\Psi_n\right]\ra_\mathrm{w},
\label{eq:FlowAngle}
\end{align}
where the POI superscript refers to particles of interest selected from a narrow transverse momentum range and the $w$ subscript means that \Fpsi{n} is averaged over the event ensemble with each event having a weight equal to the fourth power of $v_n$~\cite{Bozek:2017qir}. The double brackets refer to an average over all particles and all events, and the single brackets refer to an average over all events. A value of $\Fpsi{n}<1$ indicates the presence of \pt-dependent FAF. A large deviation from unity suggests that the symmetry plane at a specific \ptpoi deviates from the common symmetry plane. \\

The FMF are studied with
\begin{align}
\Fvn{n} &= \frac{\la\la \cos n[\varphi_1^\mathrm{POI}+\varphi_2-\varphi_3^\mathrm{POI}-\varphi_4]\ra\ra /\left(\la\la \cos n[\varphi_1^\mathrm{POI}-\varphi_3^\mathrm{a}]\ra\ra\la\la\cos n[\varphi_2-\varphi_4]\ra\ra\right)}{\la\la\cos n[\varphi_1+\varphi_2-\varphi_3-\varphi_4]\ra\ra/\la\la \cos n[\varphi_1-\varphi_2]\ra\ra^2}
=\frac{\la v_n^2(\pt)v_n^2\ra/\la v_n^2(\pt)\ra\la v_n^2\ra}{\la v_n^4\ra/\la v_n^2\ra^2}.
\label{eq:FlowMag}
\end{align}
A deviation of \Fvn{n} from unity indicates the presence of \pt-dependent FMF. The magnitude of the deviation will show how strongly the flow magnitude in a specific \pt~range, \vn(\pt), is decorrelated with respect to the integrated flow, \vn. The correlators \Fpsi{n} and \Fvn{n} probe higher moments of the distribution of flow fluctuations compared to correlators traditionally used previously with two-particle techniques~\cite{ALICE:2017lyf, CMS:2013bza, Khachatryan:2015oea}. The lower-order moments of the FAF and FMF cannot be measured separately in experiments~\cite{Heinz:2013bua,Bozek:2021zim} but can be approximated by constructing the lower and upper limits of the first moment of flow angle and magnitude fluctuations, respectively. The lower limit of the first-moment FAF $\la\cos n\left[\Psi_n(\pt)-\Psi_n\right]\ra$ is calculated with a double angle formula as
\begin{align}
\sqrt{\frac{\Fpsi{n}+1}{2}} \geq \la\cos n[\Psi_n(\pt)-\Psi_n]\ra.
\label{eq:AngleLimit}
\end{align}
The flow vector fluctuations are calculated as the ratio of the \pt-differential flow coefficient, defined as
\begin{align}
v_n\{2\} = \frac{\la\la\cos n[\varphi_1^\mathrm{POI}-\varphi_2]\ra\ra}{\la\la\cos n[\varphi_1-\varphi_2]\ra\ra} = \frac{\la v_n(\pt)v_n\cos n [\Psi_n(\pt)-\Psi_n]\ra}{\sqrt{\la v_n^2\ra}}
\end{align} 
and the \pt-integrated flow coefficient in a narrow \pt interval~\cite{Heinz:2013bua}, i.e.
\begin{align}
v_n[2] = \la\la\cos n[\varphi_1^\mathrm{POI}-\varphi_2^\mathrm{POI}]\ra\ra = \sqrt{\la v_n^2(\pt)\ra}.
\end{align}
The flow vector fluctuations is then given by
\begin{align}
    v_n\{2\}/v_n[2] = \frac{\la v_n(\pt)~v_n\cos n\left[\Psi_n(\pt)-\Psi_n\right]\ra}{\sqrt{\la v_n^2(\pt)\ra)}\sqrt{\la v_n^2\ra}},
\label{eq:vnratio}
\end{align}
which satisfies the Cauchy-Schwartz inequality for two observables X and Y, $\langle XY\rangle\leq \sqrt{\la X^2\ra\la Y^2\ra}$, as $\cos n(\Psi_n(\pt)-\Psi_n)\leq 1$. The ratio of Eq. (\ref{eq:AngleLimit}) and (\ref{eq:vnratio}) determines the upper limit of the first-order FMF
\begin{align}
\frac{v_n\{2\}/v_n[2]}{\sqrt{(\Fpsi{n}+1)/2}} \leq \frac{\la v_n(\pt)~v_n\ra}{\sqrt{\la v_n^2(\pt)\ra}\sqrt{\la v_n^2\ra}}.
\label{eq:Rmag}
\end{align}
The limits on the first-moment flow angle and magnitude fluctuations connect the study of the separated fluctuations with prior studies of flow vector fluctuations based on two-particle correlations~\cite{ALICE:2017lyf,CMS:2013bza,Khachatryan:2015oea}. All the above correlators are calculated with the generic framework~\cite{Bilandzic:2013kga,Moravcova:2020wnf}, which corrects the non-uniformities in the acceptance of the detector. The statistical uncertainties of these two correlators in Eqs. (\ref{eq:FlowAngle}) and (\ref{eq:FlowMag}) are estimated with the bootstrap method of random sampling with replacement.

The correlators \Fpsi{2} and \Fvn{2} are measured based on 54 million \PbPb collisions recorded with the ALICE experiment~\cite{ALICE:2008ngc} in 2015 at a centre-of-mass energy per nucleon pair of \snn = 5.02 TeV. Experimentally, events are selected based on a minimum bias trigger achieved by requiring a coincidence of signals in the two V0 scintillator arrays, V0A with a pseudorapidity range $2.8 < \eta < 5.1$, and V0C with a pseudorapidity range $-3.7 < \eta < -1.7$. Additionally, a reconstructed primary vertex within $\pm10$ cm of the nominal interaction point along the beam axis is required. Events with significant pileup from out-of-bunch collisions within the Time Projection Chamber (TPC) readout time will have incorrect multiplicity and cannot be used to assess the collision properties of a given centrality class. Such pileup events are rejected based on cuts of the correlation between the number of tracks measured with different detectors. A variation of the criteria for pileup rejection is considered for the systematic uncertainties~\cite{Abelev:2014ffa}. The centrality of the events is measured using information from the V0A and V0C detectors~\cite{ALICE:2013hur}. Charged particle tracks, hereafter simply called tracks, are reconstructed using the Inner Tracking System (ITS)~\cite{Aamodt:2010aa} and the TPC~\cite{Alme:2010ke}. Tracks are selected with at least 70 TPC space points out of 159 points and a $\chi^2$ per TPC cluster less than $4$~\cite{ALICE-PUBLIC-2017-005}. To reduce contamination from secondary particles, tracks are selected with a distance of closest approach to the primary vertex of less than 2 cm in the longitudinal direction and a \pt-dependent selection ranging from 0.2 cm at \pt = 0.2 GeV\textit{/c} to 0.016 cm at \pt = 5 GeV\textit{/c} in the transverse direction. Tracks are selected within the full TPC and ITS acceptance of $|\eta|<0.8$. Additionally, as the regime of hard processes is not of interest in this work, the kinematic range of the tracks is restricted to $0.2<\pt<5$ GeV\textit{/c}, where most of the tracks originate from the thermalised medium. In order to suppress non-flow correlation contributions, such as resonance decays and jets, which are not related to collective behaviour, the subevent method is utilised to calculate the correlators. In this method the event is divided into subevents separated by a gap in pseudorapidity denoted $|\Delta\eta|$. This ensures that short-range non-flow correlations between particles from the same subevent, mostly originating from resonances, are not introduced. An $\eta$-gap of $|\Delta\eta|>0.8$ is used for two-particle correlations and $|\Delta\eta|>0$ for four-particle correlations in order to ensure optimal balance between the statistical precision and non-flow suppression. To further investigate the non-flow suppression, the analysis is also performed with the so-called like-sign method, where only positively or negatively charged particle tracks are considered for analysis. The difference is less than 1\% compared to the measurements of \Fpsi{2} and\Fvn{2} using all tracks. Furthermore, the analysis is repeated with increasing pseudorapidity gaps between the subevents. Pseudorapidity gaps of $|\Delta\eta|>0$, $|\Delta\eta|>0.4$, and $\Delta\eta|>0.8$ are tested, and it is found that the measurements of \Fpsi{2} and \Fvn{2} differ by less than 1\%, when measured with different pseudorapidity gaps. Additional Monte Carlo studies using HIJING~\cite{PhysRevD.44.3501}, a model that does not feature collective effects, but involves particle correlations arising from other sources, indicate that non-flow is sufficiently suppressed with the applied pseudorapidity gaps. The HIJING calculations of the four-particle correlation functions defined in Eqs. (\ref{eq:FlowAngle}) and (\ref{eq:FlowMag}) show no statistically significant difference from zero. Based on the model studies, the like-sign method, and the variations of the pseudorapidity gaps, the non-flow correlation contributions are less than 1\% of the measured values of \Fpsi{2} and \Fvn{2}.

Systematic uncertainties are evaluated by varying the event and track selection criteria. The systematic uncertainty is presented for \Fpsi{2}, as the uncertainties are of a similar size for both observables. Since the systematic uncertainty may depend on the collision centrality and the \pt bin, only the largest contribution from each source is mentioned below. The systematic uncertainty associated with the event selection criteria is evaluated by varying the selection on the vertex position along the beam direction (from 10 cm to 7, 8, or 9 cm), the magnetic field polarity, and the criteria for rejecting pileup events, and is below 1\%. 
The robustness of the centrality determination is investigated by repeating the analysis using the centrality estimated at midrapidity from hits in the Silicon Pixel Detector (SPD) instead of the V0, resulting in a negligible difference. Uncertainties related to track selection are estimated by considering different track reconstructions and track quality selection criteria. Changing the track reconstruction to include tracks without hits in the SPD leads to a variation of, at most, $1.7\%$. The quality of the reconstructed tracks is varied by changing the minimum number of TPC space points to $80$ and $90$, which leads to a difference of $1.5\%$ on the measured correlators. Uncertainties related to the variations in the distance of closest approach in both longitudinal and transverse directions are negligible, indicating that the effect of contaminations from secondary particles on the measurements is insignificant. Finally, tightening the  $\chi^2$ per number of TPC clusters from 4 to 2.5 gives an uncertainty of, at most, $3\%$. The total systematic uncertainty is calculated as the quadratic sum of the individual sources that have a statistically significant contribution according to a statistical test~\cite{Barlow:2002yb}.

\begin{figure}[t]
    \centering
    \includegraphics[width=\textwidth]{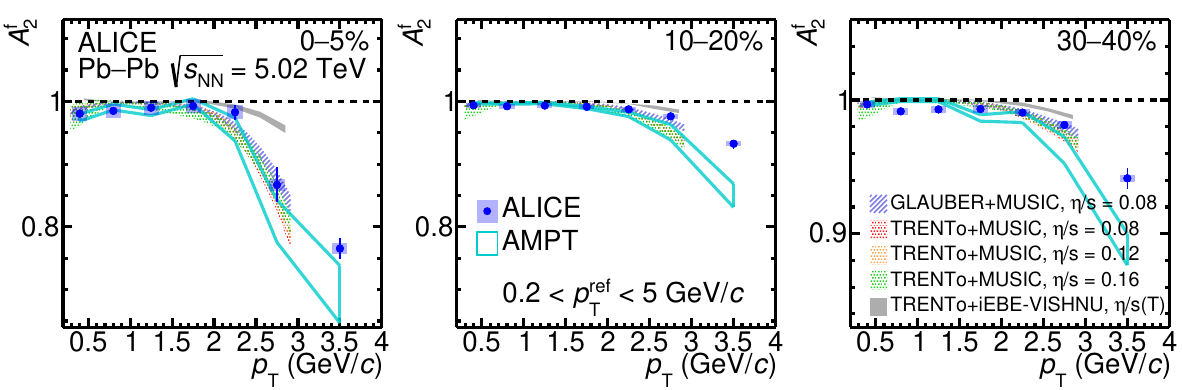}
    \caption{The flow angle fluctuation \Fpsi{2} in \PbPb collisions at \snn = 5.02 TeV as a function of the transverse momentum, \pt, in centrality classes 0--5\% (left), 10--20\% (middle), and 30--40\% (right).  Comparison with iEBE-VISHNU with \trento initial conditions and $\eta/s(T)$~\cite{Zhao:2017yhj}, MUSIC with Glauber initial conditions and $\eta/s=0.08$~\cite{Bozek:2021mov}, MUSIC with \trento initial conditions and $\eta/s=0.08,0.12,0.16$~\cite{Bozek:2021mov}, and AMPT~\cite{Ma:2016fve} are shown as coloured bands.}
    \label{fig:Fpsi}
\end{figure}

Measured values of \Fpsi{2} are present in Fig.~\ref{fig:Fpsi} as a function of the transverse momentum \pt in selected collision centrality classes. The results are presented from 0.2 GeV\textit{/c} up to 4 GeV\textit{/c} since the requirement of two particles at high \pt for the four-particle correlations limits the available data sample. In the 0--5\% most central collisions, finite and increasing large deviations from unity are observed starting from \pt $\approx$ 2.5 GeV\textit{/c}. As previously mentioned, this deviation cannot be attributed to non-flow effects, whose contributions are negligible. With more than $5\sigma$ significance of the deviation at \pt $> 3$ GeV/\textit{c} across the presented centralities, these measurements provide the first observation of \pt-dependent FAF. In centralities 10--20\% and 30--40\%, the fluctuation weakens in comparison to 0--5\% most central collisions and reaches around 5--7\% deviation from unity at $3 < \pt < 4$ GeV/\textit{c}. The increasing deviation from unity with \pt $> 2.5$ GeV\textit{/c} observed in data suggests that the elliptic flow at large transverse momentum ($\pt > 2.5$ GeV\textit{/c}) may not be correlated with the reference flow and a common symmetry plane. This will affect the comparison of measurements relying on a common symmetry plane between particles at high and low \pt with theoretical models that do not feature FAF.

Theoretical calculations with AMPT~\cite{Lin:2004en,Ma:2016fve}, MUSIC~\cite{Schenke:2010rr,Schenke:2010nt}, and iEBE-VISHNU~\cite{Shen:2014vra} models are, when available, compared to the data in Fig.~\ref{fig:Fpsi}. The AMPT transport model uses partonic interactions within the string melting tune, while a quark coalescence model is utilised to form hadrons, which are then transported through a hadronic cascade model~\cite{Li:1995pra}. The input parameters of the AMPT model are tuned to measurements of $\mathrm{d}N/\mathrm{d}\eta$, \pt-spectra and elliptic flow of charged pions, kaons and protons from ALICE ~\cite{Xu:2011fi,Ma:2016fve}. On the other hand, the MUSIC model is an event-by-event (3+1D) viscous hydrodynamic model and is used with both Glauber~\cite{Bozek:2019wyr} and \trento~\cite{Moreland:2014oya} initial conditions. Different values of $\eta/s$ of 0.08, 0.12, and 0.16 are used with \trento initial conditions~\cite{Bozek:2021mov}. Finally, the iEBE-VISHNU model is an event-by-event (2+1)D viscous hydrodynamical model coupled to the hadronic cascade model UrQMD~\cite{Bass:1998ca}. This model has been successful in describing collective phenomena, as well as event-by-event fluctuations, in several collision systems and energies~\cite{Zhao:2017yhj,Li:2021nas}. The iEBE-VISHNU model calculations with \trento initial conditions and a temperature-dependent specific shear viscosity $\eta/s(T)$~\cite{Bernhard:2016tnd} are also shown. The iEBE-VISHNU model calculations are available at \pt ranges below 3 GeV\textit{/c}. 

The AMPT calculation presented here describes the data well in the 0--5\% most central collisions and captures the deviation from unity in the highest \pt~bin. At higher centralities, the AMPT calculation overestimates the deviation from unity at high \pt. The comparison of the MUSIC calculations with Glauber and \trento initial conditions shows that \Fpsi{2} is sensitive to the fluctuations in the initial state with little to no sensitivity to the different values of specific shear viscosity. This observation is consistent with the AMPT calculations presented in~\cite{Nielsen:2022jms}, where AMPT with different values of specific shear viscosity and initial conditions are compared. The iEBE-VISHNU calculation underestimates the deviation of \Fpsi{2} from unity at \pt $> 2.5$ GeV\textit{/c} across the presented centralities with the largest difference in the 0--5\% most central collisions. The iEBE-VISHNU model with \trento initial conditions uses parameters extracted from a Bayesian analysis~\cite{Bernhard:2016tnd} in contrast to the MUSIC models, which use standard \trento initial conditions with p = 0~\cite{Moreland:2014oya}. The extracted parameters from the Bayesian analysis represent the current best understanding of the initial conditions and QGP transport properties. Including additional constraints in the Bayesian analyses, such as \Fpsi{2}, should improve our understanding of the event-by-event fluctuating initial state thus allowing for a more robust and unbiased extraction of the expansion properties of the matter formed in these collisions.

\begin{figure}[t]
    \centering
    \includegraphics[width=\textwidth]{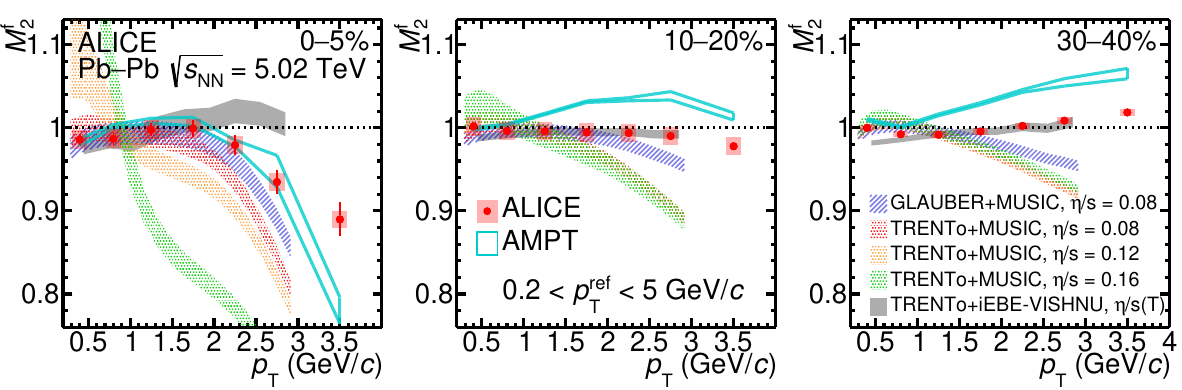}
    \caption{The flow magnitude fluctuation \Fvn{2} in \PbPb collisions at \snn = 5.02 TeV as a function of the transverse momentum, \pt, in centrality classes 0--5\% (left), 10--20\% (middle), and 30--40\% (right). Comparison with iEBE-VISHNU with \trento initial conditions and $\eta/s(T)$~\cite{Zhao:2017yhj}, MUSIC with Glauber initial conditions and $\eta/s=0.08$~\cite{Bozek:2021mov}, MUSIC with \trento initial conditions and $\eta/s=0.08,0.12,0.16$~\cite{Bozek:2021mov}, and AMPT~\cite{Ma:2016fve} are shown as coloured bands.}
    \label{fig:Fvn}
\end{figure}

\begin{figure}[t]
    \centering
    \includegraphics[scale=0.6]{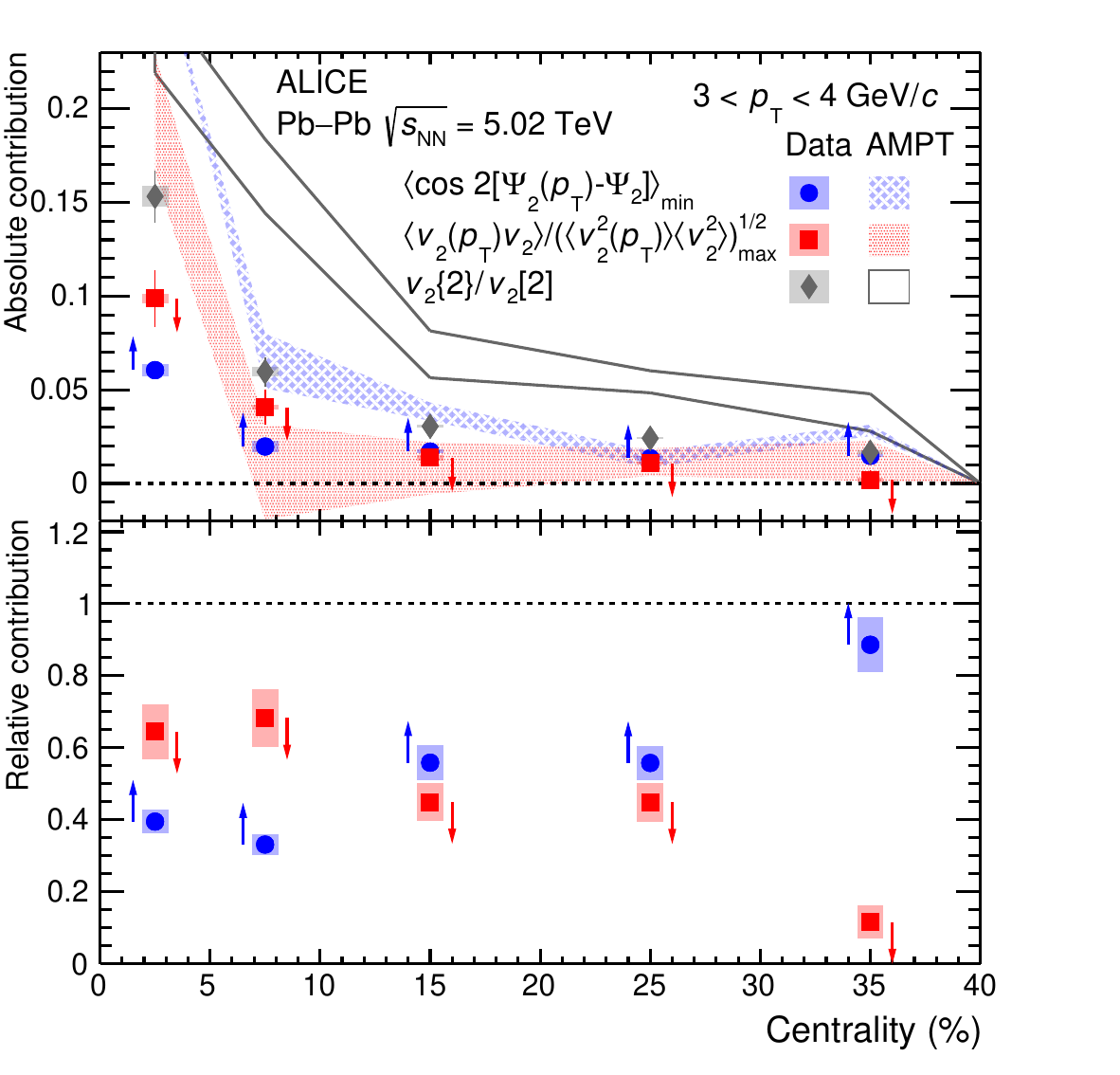}
    \caption{The lower limit of the first-order flow angle fluctuations, upper limit of the first-order flow magnitude fluctuations, and the flow vector fluctuations as a function of centrality for the $3.0<\pt<4.0$ GeV\textit{/c}. The lower and upper limits are denoted by the arrows. The top panel shows the absolute contribution and the bottom panel the contribution relative to the overall flow vector fluctuations. Comparisons to AMPT~\cite{Ma:2016fve} are shown as coloured bands.}
    \label{fig:Centdep}
\end{figure}
The measurements of the \pt-dependent FMF, \Fvn{2} in centrality classes 0--5\%, 10--20\%, and 30--40\% are shown in Fig.~\ref{fig:Fvn}. 
A substantial deviation of \Fvn{2} from unity is observed in the 0--5\% most central collisions. This deviation by more than $5\sigma$ significance at \pt $> 3$ GeV/\textit{c} constitutes the first observation of \pt-dependent FMF. The measurements show that the \pt-differential flow coefficients decorrelate with the \pt-integrated flow coefficient as the transverse momentum increases. In centrality 0--5\%, where the initial state fluctuations are most significant, the flow magnitude deviates from unity for \pt above 2 GeV\textit{/c}. Deviations increase with rising \pt~and are more pronounced in the 0--5\% central collisions compared to those observed in the 10--20\% and 30--40\% centrality ranges. \Fvn{2} is not restricted to be below unity as seen in Fig.~\ref{fig:Fvn} at 30--40\% centrality. The observable \Fvn{n} is not constructed to satisfy the Cauchy-Schwarz inequality as such an observable, $\la v_n^2(\pt)\vnsq\ra/\sqrt{v_n^4(\pt)v_n^4}$, would require four particles from a narrow \pt~range. The flow vector fluctuations measured with two-particle correlations and the FAF measured with \Fpsi{2}, however, can only be larger than unity due to non-flow effects~\cite{Gardim:2012im}. The non-flow studies mentioned previously show that non-flow correlations are negligible for \Fvn{2}, so the deviation from unity must be due to the FMF.

The AMPT transport model calculation succeeds in describing the FMF in the most central collisions, where it also describes the FAF (Fig.~\ref{fig:Fpsi} left). At higher centralities, the AMPT model significantly overestimates the data even at low \pt, but it qualitatively captures the increasing trend of \Fvn{2} in the 30--40\% centrality interval. The AMPT model works well in qualitatively describing both FAF and FMF without a hydrodynamic phase. This is possibly due to the fact that these effects mainly originate in the initial state. The MUSIC models show strong sensitivity to the $\eta/s$ in 0--5\% most central collisions as well as a sensitivity to the different initial conditions. The dependence on $\eta/s$ is unexpected and in contrast to the findings in~\cite{Nielsen:2022jms}, where the magnitude of the flow fluctuations is found to be independent of the value of $\eta/s$. In the 10--20\% and 30--40\% centrality intervals, \Fvn{2} shows no sensitivity to $\eta/s$ but is still affected by the different initial conditions. The MUSIC models overestimate the deviation of \Fvn{2} from unity in all presented centrality classes. The iEBE-VISHNU calculations underestimate the effect in 0--5\% centrality showing almost no \pt~dependence. In the 10--20\% and 30--40\% centrality intervals, the model captures the increasing trend with \pt and, consequently, describes the data. The comparison of the measurements with the above models confirms that the FMF are driven by initial state fluctuations in non-central collisions. However, the dependence of \Fvn{2} on $\eta/s$ in the 0-5\% central collisions with the MUSIC model is not well understood. The large discrepancy between the iEBE-VISHNU model and the data in central collisions suggests that the FMF could be an important observable in Bayesian analyses and could contribute further improvements of the extracted parameters used in the state-of-the-art description of the initial state and the QGP transport properties. 

Equations (\ref{eq:AngleLimit}) and (\ref{eq:Rmag}) allow for the determination of the lower and upper limit for the contribution of the FAF and FMF, respectively, to the flow vector fluctuations defined in Eq. (\ref{eq:vnratio}). Thus the lower moments of the FAF and FMF, which cannot be directly accessed in experiments, can be explored. The lower limit on the first-order FAF, the upper limit on the first-order FMF, and the total flow vector fluctuations are shown as a function of centrality for the $3 < \pt < 4$ GeV\textit{/c} range in Fig.~\ref{fig:Centdep}. In central collisions, the upper limit on the FMF is higher than the lower limit of the FAF up to 10\% centrality. For the 10--30\% centrality interval, the limits are similar, and above 30\% centrality, the flow magnitude upper limits are much smaller, and the FAF dominate the overall flow vector fluctuations completely. This is consistent with the measurements shown in Figs.~\ref{fig:Fpsi} and~\ref{fig:Fvn}, where \Fvn{2} approaches unity faster with increasing centrality than \Fpsi{2} and even goes above unity in semicentral collisions. While \pt-dependent FAF and FMF are both present in central \PbPb collisions, the effects of FAF are present in all centralities considered in this work compared to the much smaller FMF in non-central collisions. The AMPT transport model calculations overestimate the flow vector fluctuations $v_2\{2\}/v_2[2]$ as well as the limits of the first-order FAF and FMF in the central collisions. At higher centralities, AMPT describes the lower and upper limits well, whereas the flow vector fluctuations are still overestimated by the model, as in central collisions. Hydrodynamic calculations for the limits are not available. 

In summary, the \pt-dependent flow angle fluctuations and flow magnitude fluctuations of the second-order flow vector $\vec{V}_2$ are measured in \PbPb collisions at \snn = 5.02 TeV with the correlators \Fpsi{2} and \Fvn{2} in centrality intervals 0--5\%, 10--20\%, and 30--40\%. Large deviations from unity of both \Fpsi{2} ($\approx$20\%) and \Fvn{2} ($\approx$10\%) are observed at $\pt > 3$ GeV/\textit{c} in central collisions, where the event-by-event fluctuations of the position of the colliding nucleons dominate over the geometric response. In semicentral collisions, both flow angle and flow magnitude fluctuations decrease. The flow angle fluctuations reach 5--7\%, and the flow magnitude fluctuations reach around 2\% and are above unity in 30--40\% centrality for the presented \pt~range. The flow magnitude fluctuations decrease faster than the flow angle fluctuations up to 40\% centrality, where the flow vector fluctuations are almost solely due to flow angle fluctuations. The comparison of the measurements to theoretical models shows that the observables are sensitive to the initial conditions of heavy-ion collisions. This suggests that the fluctuations originate during the early stages of the QGP evolution. The observation of flow angle and flow magnitude fluctuations thus gives additional insight into the nature of event-by-event fluctuations in the initial state of heavy-ion collisions and can be used to constrain the initial conditions and QGP properties.

\newenvironment{acknowledgement}{\relax}{\relax}
\begin{acknowledgement}
\section*{Acknowledgments}

The ALICE Collaboration would like to thank all its engineers and technicians for their invaluable contributions to the construction of the experiment and the CERN accelerator teams for the outstanding performance of the LHC complex.
The ALICE Collaboration gratefully acknowledges the resources and support provided by all Grid centres and the Worldwide LHC Computing Grid (WLCG) collaboration.
The ALICE Collaboration acknowledges the following funding agencies for their support in building and running the ALICE detector:
A. I. Alikhanyan National Science Laboratory (Yerevan Physics Institute) Foundation (ANSL), State Committee of Science and World Federation of Scientists (WFS), Armenia;
Austrian Academy of Sciences, Austrian Science Fund (FWF): [M 2467-N36] and Nationalstiftung f\"{u}r Forschung, Technologie und Entwicklung, Austria;
Ministry of Communications and High Technologies, National Nuclear Research Center, Azerbaijan;
Conselho Nacional de Desenvolvimento Cient\'{\i}fico e Tecnol\'{o}gico (CNPq), Financiadora de Estudos e Projetos (Finep), Funda\c{c}\~{a}o de Amparo \`{a} Pesquisa do Estado de S\~{a}o Paulo (FAPESP) and Universidade Federal do Rio Grande do Sul (UFRGS), Brazil;
Bulgarian Ministry of Education and Science, within the National Roadmap for Research Infrastructures 2020¿2027 (object CERN), Bulgaria;
Ministry of Education of China (MOEC) , Ministry of Science \& Technology of China (MSTC) and National Natural Science Foundation of China (NSFC), China;
Ministry of Science and Education and Croatian Science Foundation, Croatia;
Centro de Aplicaciones Tecnol\'{o}gicas y Desarrollo Nuclear (CEADEN), Cubaenerg\'{\i}a, Cuba;
Ministry of Education, Youth and Sports of the Czech Republic, Czech Republic;
The Danish Council for Independent Research | Natural Sciences, the VILLUM FONDEN and Danish National Research Foundation (DNRF), Denmark;
Helsinki Institute of Physics (HIP), Finland;
Commissariat \`{a} l'Energie Atomique (CEA) and Institut National de Physique Nucl\'{e}aire et de Physique des Particules (IN2P3) and Centre National de la Recherche Scientifique (CNRS), France;
Bundesministerium f\"{u}r Bildung und Forschung (BMBF) and GSI Helmholtzzentrum f\"{u}r Schwerionenforschung GmbH, Germany;
General Secretariat for Research and Technology, Ministry of Education, Research and Religions, Greece;
National Research, Development and Innovation Office, Hungary;
Department of Atomic Energy Government of India (DAE), Department of Science and Technology, Government of India (DST), University Grants Commission, Government of India (UGC) and Council of Scientific and Industrial Research (CSIR), India;
National Research and Innovation Agency - BRIN, Indonesia;
Istituto Nazionale di Fisica Nucleare (INFN), Italy;
Japanese Ministry of Education, Culture, Sports, Science and Technology (MEXT) and Japan Society for the Promotion of Science (JSPS) KAKENHI, Japan;
Consejo Nacional de Ciencia (CONACYT) y Tecnolog\'{i}a, through Fondo de Cooperaci\'{o}n Internacional en Ciencia y Tecnolog\'{i}a (FONCICYT) and Direcci\'{o}n General de Asuntos del Personal Academico (DGAPA), Mexico;
Nederlandse Organisatie voor Wetenschappelijk Onderzoek (NWO), Netherlands;
The Research Council of Norway, Norway;
Commission on Science and Technology for Sustainable Development in the South (COMSATS), Pakistan;
Pontificia Universidad Cat\'{o}lica del Per\'{u}, Peru;
Ministry of Education and Science, National Science Centre and WUT ID-UB, Poland;
Korea Institute of Science and Technology Information and National Research Foundation of Korea (NRF), Republic of Korea;
Ministry of Education and Scientific Research, Institute of Atomic Physics, Ministry of Research and Innovation and Institute of Atomic Physics and University Politehnica of Bucharest, Romania;
Ministry of Education, Science, Research and Sport of the Slovak Republic, Slovakia;
National Research Foundation of South Africa, South Africa;
Swedish Research Council (VR) and Knut \& Alice Wallenberg Foundation (KAW), Sweden;
European Organization for Nuclear Research, Switzerland;
Suranaree University of Technology (SUT), National Science and Technology Development Agency (NSTDA), Thailand Science Research and Innovation (TSRI) and National Science, Research and Innovation Fund (NSRF), Thailand;
Turkish Energy, Nuclear and Mineral Research Agency (TENMAK), Turkey;
National Academy of  Sciences of Ukraine, Ukraine;
Science and Technology Facilities Council (STFC), United Kingdom;
National Science Foundation of the United States of America (NSF) and United States Department of Energy, Office of Nuclear Physics (DOE NP), United States of America.
In addition, individual groups or members have received support from:
Marie Sk\l{}odowska Curie, European Research Council, Strong 2020 - Horizon 2020 (grant nos. 950692, 824093, 896850), European Union;
Academy of Finland (Center of Excellence in Quark Matter) (grant nos. 346327, 346328), Finland;
Programa de Apoyos para la Superaci\'{o}n del Personal Acad\'{e}mico, UNAM, Mexico.

\end{acknowledgement}

\bibliographystyle{utphys}
\bibliography{bibliography}

\newpage
\appendix
\section{The ALICE Collaboration}
\label{app:collab}

\begin{flushleft} 
\small

S.~Acharya\,\orcidlink{0000-0002-9213-5329}\,$^{\rm 125,132}$, 
D.~Adamov\'{a}\,\orcidlink{0000-0002-0504-7428}\,$^{\rm 86}$, 
A.~Adler$^{\rm 69}$, 
G.~Aglieri Rinella\,\orcidlink{0000-0002-9611-3696}\,$^{\rm 32}$, 
M.~Agnello\,\orcidlink{0000-0002-0760-5075}\,$^{\rm 29}$, 
N.~Agrawal\,\orcidlink{0000-0003-0348-9836}\,$^{\rm 50}$, 
Z.~Ahammed\,\orcidlink{0000-0001-5241-7412}\,$^{\rm 132}$, 
S.~Ahmad\,\orcidlink{0000-0003-0497-5705}\,$^{\rm 15}$, 
S.U.~Ahn\,\orcidlink{0000-0001-8847-489X}\,$^{\rm 70}$, 
I.~Ahuja\,\orcidlink{0000-0002-4417-1392}\,$^{\rm 37}$
% ...

\section*{Affiliation Notes}

$^{\rm I}$ Deceased\\
$^{\rm II}$ Also at: Max-Planck-Institut f\"{u}r Physik, Munich, Germany\\
$^{\rm III}$ Also at: Italian National Agency for New Technologies, Energy and Sustainable Economic Development (ENEA), Bologna, Italy\\
$^{\rm IV}$ Also at: Dipartimento DET del Politecnico di Torino, Turin, Italy\\
$^{\rm V}$ Also at: Department of Applied Physics, Aligarh Muslim University, Aligarh, India\\
$^{\rm VI}$ Also at: Institute of Theoretical Physics, University of Wroclaw, Poland\\
$^{\rm VII}$ Also at: An institution covered by a cooperation agreement with CERN\\

\section*{Collaboration Institutes}

$^{1}$ A.I. Alikhanyan National Science Laboratory (Yerevan Physics Institute) Foundation, Yerevan, Armenia\\
$^{2}$ AGH University of Science and Technology, Cracow, Poland\\
$^{3}$ Bogolyubov Institute for Theoretical Physics, National Academy of Sciences of Ukraine, Kiev, Ukraine\\
$^{4}$ Bose Institute, Department of Physics  and Centre for Astroparticle Physics and Space Science (CAPSS), Kolkata, India\\
$^{5}$ California Polytechnic State University, San Luis Obispo, California, United States\\
$^{6}$ Central China Normal University, Wuhan, China\\
$^{7}$ Centro de Aplicaciones Tecnol\'{o}gicas y Desarrollo Nuclear (CEADEN), Havana, Cuba\\
$^{8}$ Centro de Investigaci\'{o}n y de Estudios Avanzados (CINVESTAV), Mexico City and M\'{e}rida, Mexico\\
$^{9}$ Chicago State University, Chicago, Illinois, United States\\
$^{10}$ China Institute of Atomic Energy, Beijing, China\\
$^{11}$ Chungbuk National University, Cheongju, Republic of Korea\\
$^{12}$ Comenius University Bratislava, Faculty of Mathematics, Physics and Informatics, Bratislava, Slovak Republic\\
$^{13}$ COMSATS University Islamabad, Islamabad, Pakistan\\
$^{14}$ Creighton University, Omaha, Nebraska, United States\\
$^{15}$ Department of Physics, Aligarh Muslim University, Aligarh, India\\
$^{16}$ Department of Physics, Pusan National University, Pusan, Republic of Korea\\
$^{17}$ Department of Physics, Sejong University, Seoul, Republic of Korea\\
$^{18}$ Department of Physics, University of California, Berkeley, California, United States\\
$^{19}$ Department of Physics, University of Oslo, Oslo, Norway\\
$^{20}$ Department of Physics and Technology, University of Bergen, Bergen, Norway\\
$^{21}$ Dipartimento di Fisica, Universit\`{a} di Pavia, Pavia, Italy\\
$^{22}$ Dipartimento di Fisica dell'Universit\`{a} and Sezione INFN, Cagliari, Italy\\
$^{23}$ Dipartimento di Fisica dell'Universit\`{a} and Sezione INFN, Trieste, Italy\\
$^{24}$ Dipartimento di Fisica dell'Universit\`{a} and Sezione INFN, Turin, Italy\\
$^{25}$ Dipartimento di Fisica e Astronomia dell'Universit\`{a} and Sezione INFN, Bologna, Italy\\
$^{26}$ Dipartimento di Fisica e Astronomia dell'Universit\`{a} and Sezione INFN, Catania, Italy\\
$^{27}$ Dipartimento di Fisica e Astronomia dell'Universit\`{a} and Sezione INFN, Padova, Italy\\
$^{28}$ Dipartimento di Fisica `E.R.~Caianiello' dell'Universit\`{a} and Gruppo Collegato INFN, Salerno, Italy\\
$^{29}$ Dipartimento DISAT del Politecnico and Sezione INFN, Turin, Italy\\
$^{30}$ Dipartimento di Scienze MIFT, Universit\`{a} di Messina, Messina, Italy\\
$^{31}$ Dipartimento Interateneo di Fisica `M.~Merlin' and Sezione INFN, Bari, Italy\\
$^{32}$ European Organization for Nuclear Research (CERN), Geneva, Switzerland\\
$^{33}$ Faculty of Electrical Engineering, Mechanical Engineering and Naval Architecture, University of Split, Split, Croatia\\
$^{34}$ Faculty of Engineering and Science, Western Norway University of Applied Sciences, Bergen, Norway\\
$^{35}$ Faculty of Nuclear Sciences and Physical Engineering, Czech Technical University in Prague, Prague, Czech Republic\\
$^{36}$ Faculty of Physics, Sofia University, Sofia, Bulgaria\\
$^{37}$ Faculty of Science, P.J.~\v{S}af\'{a}rik University, Ko\v{s}ice, Slovak Republic\\
$^{38}$ Frankfurt Institute for Advanced Studies, Johann Wolfgang Goethe-Universit\"{a}t Frankfurt, Frankfurt, Germany\\
$^{39}$ Fudan University, Shanghai, China\\
$^{40}$ Gangneung-Wonju National University, Gangneung, Republic of Korea\\
$^{41}$ Gauhati University, Department of Physics, Guwahati, India\\
$^{42}$ Helmholtz-Institut f\"{u}r Strahlen- und Kernphysik, Rheinische Friedrich-Wilhelms-Universit\"{a}t Bonn, Bonn, Germany\\
$^{43}$ Helsinki Institute of Physics (HIP), Helsinki, Finland\\
$^{44}$ High Energy Physics Group,  Universidad Aut\'{o}noma de Puebla, Puebla, Mexico\\
$^{45}$ Horia Hulubei National Institute of Physics and Nuclear Engineering, Bucharest, Romania\\
$^{46}$ Indian Institute of Technology Bombay (IIT), Mumbai, India\\
$^{47}$ Indian Institute of Technology Indore, Indore, India\\
$^{48}$ INFN, Laboratori Nazionali di Frascati, Frascati, Italy\\
$^{49}$ INFN, Sezione di Bari, Bari, Italy\\
$^{50}$ INFN, Sezione di Bologna, Bologna, Italy\\
$^{51}$ INFN, Sezione di Cagliari, Cagliari, Italy\\
$^{52}$ INFN, Sezione di Catania, Catania, Italy\\
$^{53}$ INFN, Sezione di Padova, Padova, Italy\\
$^{54}$ INFN, Sezione di Pavia, Pavia, Italy\\
$^{55}$ INFN, Sezione di Torino, Turin, Italy\\
$^{56}$ INFN, Sezione di Trieste, Trieste, Italy\\
$^{57}$ Inha University, Incheon, Republic of Korea\\
$^{58}$ Institute for Gravitational and Subatomic Physics (GRASP), Utrecht University/Nikhef, Utrecht, Netherlands\\
$^{59}$ Institute of Experimental Physics, Slovak Academy of Sciences, Ko\v{s}ice, Slovak Republic\\
$^{60}$ Institute of Physics, Homi Bhabha National Institute, Bhubaneswar, India\\
$^{61}$ Institute of Physics of the Czech Academy of Sciences, Prague, Czech Republic\\
$^{62}$ Institute of Space Science (ISS), Bucharest, Romania\\
$^{63}$ Institut f\"{u}r Kernphysik, Johann Wolfgang Goethe-Universit\"{a}t Frankfurt, Frankfurt, Germany\\
$^{64}$ Instituto de Ciencias Nucleares, Universidad Nacional Aut\'{o}noma de M\'{e}xico, Mexico City, Mexico\\
$^{65}$ Instituto de F\'{i}sica, Universidade Federal do Rio Grande do Sul (UFRGS), Porto Alegre, Brazil\\
$^{66}$ Instituto de F\'{\i}sica, Universidad Nacional Aut\'{o}noma de M\'{e}xico, Mexico City, Mexico\\
$^{67}$ iThemba LABS, National Research Foundation, Somerset West, South Africa\\
$^{68}$ Jeonbuk National University, Jeonju, Republic of Korea\\
$^{69}$ Johann-Wolfgang-Goethe Universit\"{a}t Frankfurt Institut f\"{u}r Informatik, Fachbereich Informatik und Mathematik, Frankfurt, Germany\\
$^{70}$ Korea Institute of Science and Technology Information, Daejeon, Republic of Korea\\
$^{71}$ KTO Karatay University, Konya, Turkey\\
$^{72}$ Laboratoire de Physique des 2 Infinis, Ir\`{e}ne Joliot-Curie, Orsay, France\\
$^{73}$ Laboratoire de Physique Subatomique et de Cosmologie, Universit\'{e} Grenoble-Alpes, CNRS-IN2P3, Grenoble, France\\
$^{74}$ Lawrence Berkeley National Laboratory, Berkeley, California, United States\\
$^{75}$ Lund University Department of Physics, Division of Particle Physics, Lund, Sweden\\
$^{76}$ Nagasaki Institute of Applied Science, Nagasaki, Japan\\
$^{77}$ Nara Women{'}s University (NWU), Nara, Japan\\
$^{78}$ National and Kapodistrian University of Athens, School of Science, Department of Physics , Athens, Greece\\
$^{79}$ National Centre for Nuclear Research, Warsaw, Poland\\
$^{80}$ National Institute of Science Education and Research, Homi Bhabha National Institute, Jatni, India\\
$^{81}$ National Nuclear Research Center, Baku, Azerbaijan\\
$^{82}$ National Research and Innovation Agency - BRIN, Jakarta, Indonesia\\
$^{83}$ Niels Bohr Institute, University of Copenhagen, Copenhagen, Denmark\\
$^{84}$ Nikhef, National institute for subatomic physics, Amsterdam, Netherlands\\
$^{85}$ Nuclear Physics Group, STFC Daresbury Laboratory, Daresbury, United Kingdom\\
$^{86}$ Nuclear Physics Institute of the Czech Academy of Sciences, Husinec-\v{R}e\v{z}, Czech Republic\\
$^{87}$ Oak Ridge National Laboratory, Oak Ridge, Tennessee, United States\\
$^{88}$ Ohio State University, Columbus, Ohio, United States\\
$^{89}$ Physics department, Faculty of science, University of Zagreb, Zagreb, Croatia\\
$^{90}$ Physics Department, Panjab University, Chandigarh, India\\
$^{91}$ Physics Department, University of Jammu, Jammu, India\\
$^{92}$ Physics Department, University of Rajasthan, Jaipur, India\\
$^{93}$ Physics Program and International Institute for Sustainability with Knotted Chiral Meta Matter (SKCM2), Hiroshima University, Hiroshima, Japan\\
$^{94}$ Physikalisches Institut, Eberhard-Karls-Universit\"{a}t T\"{u}bingen, T\"{u}bingen, Germany\\
$^{95}$ Physikalisches Institut, Ruprecht-Karls-Universit\"{a}t Heidelberg, Heidelberg, Germany\\
$^{96}$ Physik Department, Technische Universit\"{a}t M\"{u}nchen, Munich, Germany\\
$^{97}$ Politecnico di Bari and Sezione INFN, Bari, Italy\\
$^{98}$ Research Division and ExtreMe Matter Institute EMMI, GSI Helmholtzzentrum f\"ur Schwerionenforschung GmbH, Darmstadt, Germany\\
$^{99}$ Saga University, Saga, Japan\\
$^{100}$ Saha Institute of Nuclear Physics, Homi Bhabha National Institute, Kolkata, India\\
$^{101}$ School of Physics and Astronomy, University of Birmingham, Birmingham, United Kingdom\\
$^{102}$ Secci\'{o}n F\'{\i}sica, Departamento de Ciencias, Pontificia Universidad Cat\'{o}lica del Per\'{u}, Lima, Peru\\
$^{103}$ Stefan Meyer Institut f\"{u}r Subatomare Physik (SMI), Vienna, Austria\\
$^{104}$ SUBATECH, IMT Atlantique, Nantes Universit\'{e}, CNRS-IN2P3, Nantes, France\\
$^{105}$ Suranaree University of Technology, Nakhon Ratchasima, Thailand\\
$^{106}$ Technical University of Ko\v{s}ice, Ko\v{s}ice, Slovak Republic\\
$^{107}$ The Henryk Niewodniczanski Institute of Nuclear Physics, Polish Academy of Sciences, Cracow, Poland\\
$^{108}$ The University of Texas at Austin, Austin, Texas, United States\\
$^{109}$ Universidad Aut\'{o}noma de Sinaloa, Culiac\'{a}n, Mexico\\
$^{110}$ Universidade de S\~{a}o Paulo (USP), S\~{a}o Paulo, Brazil\\
$^{111}$ Universidade Estadual de Campinas (UNICAMP), Campinas, Brazil\\
$^{112}$ Universidade Federal do ABC, Santo Andre, Brazil\\
$^{113}$ University of Cape Town, Cape Town, South Africa\\
$^{114}$ University of Houston, Houston, Texas, United States\\
$^{115}$ University of Jyv\"{a}skyl\"{a}, Jyv\"{a}skyl\"{a}, Finland\\
$^{116}$ University of Kansas, Lawrence, Kansas, United States\\
$^{117}$ University of Liverpool, Liverpool, United Kingdom\\
$^{118}$ University of Science and Technology of China, Hefei, China\\
$^{119}$ University of South-Eastern Norway, Kongsberg, Norway\\
$^{120}$ University of Tennessee, Knoxville, Tennessee, United States\\
$^{121}$ University of the Witwatersrand, Johannesburg, South Africa\\
$^{122}$ University of Tokyo, Tokyo, Japan\\
$^{123}$ University of Tsukuba, Tsukuba, Japan\\
$^{124}$ University Politehnica of Bucharest, Bucharest, Romania\\
$^{125}$ Universit\'{e} Clermont Auvergne, CNRS/IN2P3, LPC, Clermont-Ferrand, France\\
$^{126}$ Universit\'{e} de Lyon, CNRS/IN2P3, Institut de Physique des 2 Infinis de Lyon, Lyon, France\\
$^{127}$ Universit\'{e} de Strasbourg, CNRS, IPHC UMR 7178, F-67000 Strasbourg, France, Strasbourg, France\\
$^{128}$ Universit\'{e} Paris-Saclay Centre d'Etudes de Saclay (CEA), IRFU, D\'{e}partment de Physique Nucl\'{e}aire (DPhN), Saclay, France\\
$^{129}$ Universit\`{a} degli Studi di Foggia, Foggia, Italy\\
$^{130}$ Universit\`{a} del Piemonte Orientale, Vercelli, Italy\\
$^{131}$ Universit\`{a} di Brescia, Brescia, Italy\\
$^{132}$ Variable Energy Cyclotron Centre, Homi Bhabha National Institute, Kolkata, India\\
$^{133}$ Warsaw University of Technology, Warsaw, Poland\\
$^{134}$ Wayne State University, Detroit, Michigan, United States\\
$^{135}$ Westf\"{a}lische Wilhelms-Universit\"{a}t M\"{u}nster, Institut f\"{u}r Kernphysik, M\"{u}nster, Germany\\
$^{136}$ Wigner Research Centre for Physics, Budapest, Hungary\\
$^{137}$ Yale University, New Haven, Connecticut, United States\\
$^{138}$ Yonsei University, Seoul, Republic of Korea\\
$^{139}$  Zentrum  f\"{u}r Technologie und Transfer (ZTT), Worms, Germany\\
$^{140}$ Affiliated with an institute covered by a cooperation agreement with CERN\\
$^{141}$ Affiliated with an international laboratory covered by a cooperation agreement with CERN.\\

\end{flushleft}


\end{document}